\documentclass[reprint,superscriptaddress,nofootinbib]{revtex4-1}

\usepackage{graphicx}
\usepackage{dcolumn}
\usepackage{bm}
\usepackage{float}
\usepackage{amsthm,amssymb,amsmath,url,braket,amsbsy} 
\usepackage{threeparttable}
\usepackage[utf8]{inputenc}
\graphicspath {{./figure/}}
\usepackage{color}
\usepackage{cancel}

\begin{document}

\title{Inverse Faraday effect in massive Dirac electrons}

\author{Guanxiong Qu}
\affiliation{RIKEN Center for Emergent Matter Science (CEMS), Wako 351-0198, Japan.}
\author{Gen Tatara}
\affiliation{RIKEN Center for Emergent Matter Science (CEMS), Wako 351-0198, Japan.}

\date{\today}

\begin{abstract}
We study the inverse Faraday effect (IFE) in a Dirac Hamiltonian with random impurities using Keldysh formalism and diagrammatic perturbation theory. The mass term in the Dirac Hamiltonian is essential for IFE, where the spin magnetic moment induced by circularly polarized light is proportional to the frequency of the incident light within the THz regime. For massive Dirac electrons, the corrections due to short-range impurities on spin magnetic moment vertex exhibits mixing of the spin magnetic moment vertex and spin angular momentum vertex. The spin magnetic density response is divergently enhanced by the vertex corrections near the band edge, indicting a long-range diffusion of spin density profile in massive Dirac electrons.
\end{abstract}


\maketitle

\date{\today}

\section{Introduction}
The Dirac equation\cite{dirac1928quantum} successfully describes the dynamics of relativistic electrons. Despite its background in high energy physics, the Dirac Hamiltonian and its variations are widely exploited as low-energy effective Hamiltonians in condensed matter physics, predominantly attributed to its inherent spin(pseudo-spin) orbit interaction. Among its variations, Dirac and Weyl semimetals\cite{burkov2011weyl,yang2014classification,armitage2018weyl} which exhibit gapless excitations have drawn much attentions recently\cite{brahlek2012topological,wu2013sudden,liu2014discovery,liu2014stable}. The massive Dirac Hamiltonian also describes various gapped systems with quasi-linear band dispersion, e.g., the $L$-point of the Bismuth\cite{fuseya2009interband,fuseya2012spin,fujimoto2014transport,fuseya2015transport,fukazawa2017intrinsic}.

The inherent spin orbit coupling (SOC) enables the Dirac electrons to exhibit various spin-related responses to the external electric field. For example, the ferromagnetic Dirac Hamiltonian\cite{crepieux2001theory,fujimoto2014transport} was employed to study the theory of the anomalous Hall effect. The massive Dirac Hamiltonian has demonstrated a strong spin Hall effect inside the mass gap\cite{fuseya2012spin,fukazawa2017intrinsic}. In addition to dc electric field responses, the optical responses of Dirac electrons have also been studied theoretically, especially focusing on chirality dependent phenomena, e.g., the inverse Faraday effect (IFE)\cite{tse2010giant,misawa2011electromagnetic,taguchi2016photovoltaic,PhysRevB.94.144432,PhysRevLett.117.137203}  and circular photogalvanic effect (CPGE) \cite{hosur2011circular,taguchi2016photovoltaic} . Nevertheless, theoretical investigation has been restricted to the massless case of Dirac systems. 

The inverse Faraday effect \cite{pitaevskii1961electric,pershan1963nonlinear} phenomenologically describes the static magnetization induced by circularly polarized light. The symmetry argument\cite{landau2013electrodynamics} gives a qualitative expression of the induced effective magnetic field: $\bm{M}_{\text{eff}} \sim i \bm{\mathcal{E}}\times \bm{\mathcal{E}}^*$ where $\bm{\mathcal{E}}$ is the complex electric field. In experiments, IFE is able to reverse the magnetization of magnets with strong laser pulse\cite{kimel2005ultrafast,kirilyuk2010ultrafast}, providing an optical method of ultrafast magnetization manipulation. As was demonstrated recently, sensitive detection of the effective magnetic field induced by a continuous laser can be carried out electrically through the inverse spin Hall effect\cite{kawaguchi2020giant}. It is, thus, of interest to investigate the IFE in a massive Dirac electron's system, which was demonstrated to exhibit a large spin Hall effect \cite{fuseya2012spin}.

The Dirac Hamiltonian is also featured by its particle-hole symmetry. The negative energy states correspond to electron vacancies (holes), and thus physical operators are antisymmetric with respect to energy. However, it turns out that physical operators are mixed by interactions with their unphysical counterparts which is symmetric with energy. We shall demonstrate that even an energy-conserving impurity scattering leads to an entanglement of physical magnetic moment and unphysical spin (angular) vertices\cite{ramana1981theory}, resulting in a significant enhancement of magnetic moment response at the band edge. The entanglement is a result of an interference between positive and negative energy states like the \textit{Zitterbewegung}\cite{sakurai2006advanced} and is unique feature of Dirac electrons. Further, the IFE is known to be dissipative and extrinsic\cite{taguchi2011theory,misawa2011electromagnetic,taguchi2016photovoltaic}, wherefore the IFE of massive Dirac electrons  is also a nice paradigm to study the vertex corrections of the Dirac Hamiltonian. 

In this paper, we study the spin responses under circularly polarized light in massive Dirac electrons by using the Keldysh Green's function formalism. The spin magnetic density correlates with the chirality of the incident light by calculating the second order perturbation of gauge coupling. The response function holds for the general frequency of incident light, while the low-frequency expansion is employed for analytical results.  The electrons' lifetime and the ladder-type vertex corrections (VCs) are introduced by the short-range impurities. For impurity correction on the spin magnetic vertex, mixing of the spin vertices in the massive scenario is investigated in detail through concerning long-range diffusion. 

\section{Massive Dirac electrons}
We start from the effective Hamiltonian describing Dirac electrons:
\begin{equation}
\mathcal{H}_0= \hbar v k_i \rho_1 \otimes \sigma^i + m \rho_3 \otimes \sigma^0 ,
\label{eq:2_1}
\end{equation}
where $v$ is the Fermi velocity and $m$ is a mass term corresponding to half of the band gap\cite{fuseya2015transport}. The $\rho_\mu, \sigma^\nu$ are Pauli matrices spanning the particle-hole and spin space respectively $(\mu,\nu=0,1,2,3)$. Note that the $\rho_0,\sigma^0$ are identity matrices. $\hbar$ is the reduced Planck's constant and the repeated indices indicate summation. The band energy is $ \varepsilon_{\eta,\bm{k}} = \eta \varepsilon_{\bm{k}}  $ with $\eta$ representing the positive ($+1$) and negative ($+1$) energy bands and $ \varepsilon_{\bm{k}} \equiv  \sqrt{\hbar^2 v^2 k^2 + m^2}$.

The impurity scattering is assumed to be random and short-ranged. The impurity potential is 
\begin{align}
V_{\text{imp}} = n_{\text{i}} u   \rho_0 \otimes \sigma^0 ,
\label{eq:2_2}
\end{align}
where $n_{\text{i}}$ is impurity density and $u$ is the strength of $\delta$-function impurities. With the Born approximation, the retarded self energy\cite{fujimoto2014transport,fukazawa2017intrinsic} is 
\begin{align}
\text{Im} \Sigma^R ( {\varepsilon} ) 
& \equiv - \gamma_0 (\varepsilon )  \rho_0 \otimes \sigma^0 - \gamma_3 (\varepsilon )  \rho_3 \otimes \sigma^0 ,
\label{eq:2_3} 
\end{align}
where $\gamma_0$ and $\gamma_3$ are
\begin{subequations}
\begin{align}
\gamma_0 (\varepsilon)&= \frac{\pi}{2} n_i u^2 \nu (\varepsilon + \mu)  , \\
\gamma_3 (\varepsilon)&= \frac{\pi}2 n_i u^2 \frac{ m}{ \varepsilon +\mu } \nu (\varepsilon + \mu),
\label{eq:2_4}
\end{align}
\end{subequations}
in which the density of states is $\nu (\varepsilon + \mu) \equiv 1/V \sum\limits_{\eta, \bm{k}} \delta (\varepsilon+ \mu - \eta \varepsilon_{\bm{k} } )$. $V,\mu$ are the volume of the system and chemical potential, respectively. Note that $\text{Re}, \text{Im}$ denote the real and imaginary components.  The density of states of Dirac Hamiltonian is asymptotically proportional to the square of the energy, $\nu(\varepsilon) \propto |\varepsilon|\sqrt{\varepsilon^2 -m^2}$. 

In massive Dirac electrons, $\gamma_0$ is the damping coefficient  symmetric with respect to the positive and negative energy bands, while $\gamma_3$ is the anti-symmetric damping coefficient associated with the mass term. Including the self-energy, the retarded Green's function is $\mathcal{G}_{\bm{k}}^{R} (\omega) = [\hbar \omega + \mu - \mathcal{H}_0 -V_{imp}  - i \text{Im} \Sigma^R ( \hbar \omega ) ]^{-1}$. Note that $\varepsilon=\hbar \omega$ is the energy measured from the chemical potential $\mu$. Although the self-energy is generally energy-dependent, in the following discussion we focus on the scattering effect predominantly at the Fermi surface $\varepsilon=0$, $\mu = \eta \epsilon_{\bm{k}}$.  With on-shell condition ($\mu +\varepsilon = \eta \epsilon_{\bm{k}}$)\cite{fujimoto2021seebeck}, the damping coefficients are approximated as constants in energy and have the following relationship: $\Gamma= \gamma_0 + m/\mu \gamma_3$. Correspondingly, the electrons' lifetime in on-shell condition is defined as
\begin{align}
 \tau = \frac{\hbar}{2 \Gamma} = \frac{\hbar}{\pi n_{\text{i}} u^2 \nu(\mu)} \frac{\mu^2}{\mu^2 +m^2} .
\label{eq:2_5}
\end{align}
Note that the electrons' lifetime diverges at the band edges $\mu = \pm m$, due to the vanishing of states.
\section{Inverse Faraday effect}
The inverse Faraday effect is the nonlinear response of the electrical field. The incident light is described by velocity gauge\cite{PhysRevB.96.035431} coupling:
\begin{align}
H_{\text{em}} (\bm{x}, t) = -e  A_i (\bm{x}, t) v_i ,
\label{eq:2_6}
\end{align}
where $\bm{A} (\bm{x}, t) = \text{Re} \bm{\mathcal{A}} e^{i (\bm{q} \cdot \bm{x} -  \Omega t)} $ is the vector potential of incident light and  $ v_i \equiv   v \rho_1 \otimes \sigma^i $ is the velocity operator. The gauge field is defined by the physical external electric field: $\bm{\mathcal{A}}= -i \bm{\mathcal{E}}/\Omega$. $e$ is the elementary charge.

\subsection{Spin magnetic density response}
The induced spin magnetic density is calculated through Keldysh formalism. Thus, the expectation value of spin magnetic density is expressed by an equal space-time lesser Green's function:
\begin{align}
\braket{m^k} = - i \hbar \text{Tr} \left[ m^k G^{<} (\bm{x}, t; \bm{x},t) \right],
\label{eq:2_7}
\end{align}
where the spin magnetic operator is $m^k \equiv \rho_3 \otimes \sigma^k$ and its prefactor $- g^* \mu_B /2$\cite{fuseya2015transport} is neglected for simplicity where $g^*$ is an effective $g$-factor and $\mu_B$ is the Bohr magneton. Note that $G^{<}$ is the Green's function containing the gauge field $\bm{\mathcal{A}}$ which is treated in perturbative expansion later. 

\begin{figure}[h]
 \centering
 \includegraphics[width=8cm]{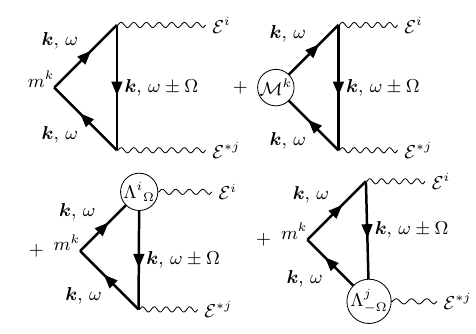}
\caption{\label{fig:1} Diagrammatic representation of the spin magnetic density induced by the circularly polarized light. The first diagram is spin magnetic response without VCs for impurity and the rest are corrections on the spin magnetic vertex and velocity vertices. The shaded circle represents the corrected vertices.} 
 \end{figure}

Since the IFE is qualitatively proportional to the cross product of the external electric field\cite{landau2013electrodynamics} $(\bm{M}_{\text{eff}} \sim i \bm{\mathcal{E}}\times \bm{\mathcal{E}}^*)$, it can be traced from the second order perturbation of the gauge coupling in Eq.~(\ref{eq:2_7}) whose corresponding diagram is shown in Fig.~\ref{fig:1}. Note that we only keep the stationary response ($0\Omega$) of the spin magnetic moment and the oscillatory response ($2\Omega$) in second-order perturbation is neglected due to its oscillation in time\cite{hertel2006theory}.
 
The spin magnetic density induced by the second-order perturbation (see Fig.~\ref{fig:1}) of the gauge field reads
\begin{align}
\braket{m^{k}}^{(2)}
&= - i  \left[  \chi^k_{ij}(+ \Omega) + \chi^k_{ji}(- \Omega)\right]  \mathcal{E}_i  \mathcal{E}_j^* ,
\label{eq:2_8}
\end{align}
where the spin magnetic response function is defined as
\begin{align}
\chi^k_{ij}( \Omega)& \equiv \frac{ \hbar  e^2 }{4 \Omega^2 V} \sum_{\bm{k},\omega} \text{Tr} \left[ m^k \mathcal{G}_{\bm{k}} (\omega)   v_i \mathcal{G}_{\bm{k}} (\omega + \Omega)    v_j  \mathcal{G}_{\bm{k}} (\omega) 
\right]^{<} .
\label{eq:2_9}
\end{align}
In thermal equilibrium, the lesser Green's function is given by $ \mathcal{G}^<_{\bm{k}} (\omega) = f  (\omega) \left[  \mathcal{G}^A_{\bm{k}} (\omega) -  \mathcal{G}^R_{\bm{k}} (\omega)  \right]$ where $ f  (\omega) $ is Fermi distribution function. 

Utilizing the spherical symmetry of the Dirac Hamiltonian and evaluating the trace term in Eq.~(\ref{eq:2_9}), it is evidenced that the response function is totally anti-symmetric $\chi^k_{ij}( \Omega) \propto \epsilon_{ijk}$, corresponding to the anti-symmetry with respect to the chirality of the incident light. The spin magnetic response in Eq.~(\ref{eq:2_8}) can be rewritten as
\begin{align}
\chi^k_{ij}(- \Omega) + \chi^k_{ji}(+ \Omega) =  \chi^k_{ji}(+ \Omega) -  \chi^k_{ji}( -\Omega).
\label{eq:2_10}
\end{align}
Apparently, only terms which are odd with external frequency $\Omega$ contribute to the IFE. To obtain an analytical expression, the response function $ \chi^k_{ji}$ is expanded in the limit $\Omega \tau \ll1$. The first order of $\Omega$-expansion inside the lesser Green's function contains the $\Omega^{-1}$ term:
\begin{widetext}
\begin{align}
\chi^{k,(-1)}_{ij}( \Omega) &= \frac{e^2 }{4 \hbar \Omega V} \sum_{\bm{k},\omega} \text{Tr} \Big[ 
  f'(\omega  ) \Big(   m^k \mathcal{G}^R_{\bm{k}} (\omega)     v_i \mathcal{G}^A_{\bm{k}} (\omega )   v_j    \mathcal{G}^A_{\bm{k}} (\omega) -   m^k\mathcal{G}^R_{\bm{k}} (\omega)   v_i \mathcal{G}^R_{\bm{k}} (\omega ) v_j     \mathcal{G}^A_{\bm{k}} (\omega) 
\Big) \notag\\
&-  f(\omega) \Big(  m^k  \mathcal{G}^A_{\bm{k}} (\omega)   v_i [ \mathcal{G}^A_{\bm{k}} (\omega) ]^2     v_j   \mathcal{G}^A_{\bm{k}} (\omega) 
- m^k \mathcal{G}^R_{\bm{k}} (\omega)    v_i  [ \mathcal{G}^R_{\bm{k}} (\omega) ]^2    v_j   \mathcal{G}^R_{\bm{k}} (\omega)   \Big)  \Big] .
\label{eq:2_11}
\end{align}
It is easily checked that the Fermi sea term $[\propto f(\omega)]$ vanishes after taking the trace and the Fermi surface term $[\propto  f'(\omega)]$
vanishes at the boundary due to the spherical symmetry of the Dirac Hamiltonian. Thus, the $\Omega^{-1}$ term totally vanishes $\chi^{k,(-1)}_{ij}( \Omega)=0$.

The third order expansion contain the $\Omega^{1}$ term:
\begin{align}
\chi^{k,(1)}_{ij}( \Omega) 
&=\frac{e^2  \Omega }{4 \hbar V}\text{Tr} \sum_{\bm{k},\omega} \Big\{  - \frac{\hbar^2}{2}  f'(\omega  )  \Big[   m^k [ \mathcal{G}^R_{\bm{k}} (\omega)  ]^2   v_i [ \mathcal{G}^A_{\bm{k}} (\omega ) ]^2  v_j    \mathcal{G}^A_{\bm{k}} (\omega)
+ m^k  \mathcal{G}^R_{\bm{k}} (\omega)     v_i [ \mathcal{G}^A_{\bm{k}} (\omega ) ]^2  v_j   [ \mathcal{G}^A_{\bm{k}} (\omega) ]^2 \notag \\
& -  m^k [ \mathcal{G}^R_{\bm{k}} (\omega)  ]^2   v_i [ \mathcal{G}^R_{\bm{k}} (\omega ) ]^2  v_j    \mathcal{G}^A_{\bm{k}} (\omega) - m^k  \mathcal{G}^R_{\bm{k}} (\omega)     v_i [ \mathcal{G}^R_{\bm{k}} (\omega ) ]^2  v_j   [ \mathcal{G}^A_{\bm{k}} (\omega) ]^2 
\Big] \notag \\
& - \hbar^3 f(\omega) \Big[ m^k  \mathcal{G}^A_{\bm{k}} (\omega)   v_i [ \mathcal{G}^A_{\bm{k}} (\omega) ]^4     v_j   \mathcal{G}^A_{\bm{k}} (\omega) 
- m^k \mathcal{G}^R_{\bm{k}} (\omega)    v_i  [ \mathcal{G}^R_{\bm{k}} (\omega) ]^4    v_j   \mathcal{G}^R_{\bm{k}} (\omega)  \Big] \Big\},
\label{eq:2_12}
\end{align}
\end{widetext}
where the Fermi sea contribution contains only retarded or advanced Green's functions which are of higher order $(\mathcal{O} (\Gamma^1)]$ than the Fermi surface term under the condition $\Gamma/\mu \ll 1$. The dominance of the Fermi surface contributions also justifies our assumption that the self-energy is confined within a small region near the Fermi surface. 
 In the zero-temperature limit, the spin magnetic density response function $[\chi^{k}_{ij}( \Omega)=\chi^{k,(1)}_{ij}( \Omega)]$ without VCs is
\begin{align}
\chi^{k}_{ij}( \Omega) 
&=  \epsilon_{ijk}   \hbar v^2 e^2    \frac{    m  ( \mu^2 -m^2) ( 2\mu^2 +3 m^2) }{12  \mu^3 (\mu^2+m^2)^2 }  \nu(\mu) \Omega \tau^2 .
\label{eq:2_13}
\end{align}
The calculation details are shown in Appendix.~\ref{Appx.A}.
The result [Eq.~(\ref{eq:2_13})] indicates that IFE vanishes for the massless $(m=0)$ case. Similarly, for the massless Weyl semimetal, the photovoltaic chiral magnetic effect can only be induced with unbalanced chemical potential\cite{taguchi2016photovoltaic}. The induced spin magnetic density changes sign for electron and hole states. Despite the divergence of the lifetime $\tau$ at the band edge $(\mu=\pm m)$, the response function $\chi^{k}_{ij}$ is proportional to $ (\mu^2 - m^2)^{1/2}$ and, thus, still vanishes at the band edge.

\subsection{Spin angular density response}
In the Dirac Hamiltonian, spin can also couple with the electron and hole states equally via the spin angular operator $(s^k \equiv \rho_0 \otimes \sigma^k)$\cite{PhysRevB.105.214419}. Note that the spin angular operator is not associated with the physical spin density but is essential to the impurity vertex corrections (see Sec.~\ref{Sec.3-3}).  By simply replacing the spin magnetic operator with the spin angular operator in Eq.~(\ref{eq:2_9}), we have the spin angular response function:
\begin{align}
\Pi^{k}_{ij}( \Omega) &= \frac{ \hbar  e^2 }{4 \Omega^2 V} \sum_{\bm{k},\omega} \text{Tr} \left[ s^k \mathcal{G}_{\bm{k}} (\omega)   v_i \mathcal{G}_{\bm{k}} (\omega + \Omega)    v_j  \mathcal{G}_{\bm{k}} (\omega)  \right]^{<} \notag \\
&=  \epsilon_{ijk} \hbar v^2 e^2 \frac{  (m^2 - \mu^2 )( \mu^2 + 2 m^2) }{ 12  \mu^2 (\mu^2 +m^2)^2  } \nu(\mu) \Omega \tau^2,
\label{eq:2_14}
\end{align}
where we similarly trace the second order perturbation with the gauge field coupling and take out the physical contribution linear with $\Omega$.
The spin angular response function, instead, is an even function with the chemical potential corresponding to the same responses in the positive and negative energy bands and does not vanish in the massless limit. 

\subsection{Impurity correction on spin magnetic and velocity vertices}\label{Sec.3-3}
 \begin{figure}[h]
 \centering
 \includegraphics[width=8cm]{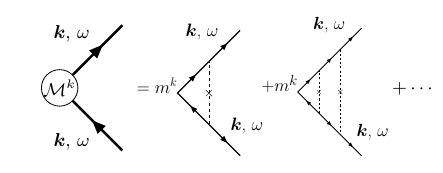} 
\caption{\label{fig:3} Diagrammatic representation the VCs ladder of spin magnetic vertex.} 
 \end{figure}
The VCs on the spin magnetic vertex and velocity vertex, represented by the infinite sum of the ladder diagrams (see Fig.~\ref{fig:3}), need to be taken into account for a consistency with inclusion of the self-energy. In the Dirac Hamiltonian, corrections on the spin magnetic vertex and velocity vertex not only iterate with themselves, but couple with its reciprocal vertices. For example, the first-order correction of the spin magnetic vertex $m^k$ contains two separate vertices, i.e., spin magnetic and spin angular vertices,
\begin{align}
\mathcal{M}^{k,(1)}_{\omega}  &= \frac{n_\text{i} u^2}{V} \sum_{\bm{k}} \mathcal{G}^R_{\bm{k}} (\omega)
m^k \mathcal{G}^A_{\bm{k}} (\omega) \notag \\
&= A^{\mathcal{M}}(\omega) s^k  + S^{\mathcal{M}}(\omega) m^k,
\label{eq:2_15}
\end{align}
where the symmetric $S^{\mathcal{M}}(\omega)$ and anti-symmetric $A^{\mathcal{M}}(\omega)$ coefficients are 
\begin{subequations}
\begin{align}
\label{eq:2_16a}
S^{\mathcal{M}}(\omega) &= \frac{ \frac{2}{3} (\hbar \omega + \mu)^2 +\frac{1}{3}m^2 }{ (\hbar \omega +\mu)^2 +m^2} + \mathcal{O} (\Gamma^2), \\
A^{\mathcal{M}}(\omega) &=    \frac{(\hbar \omega +\mu)m}{ (\hbar \omega +\mu)^2 +m^2}  + \mathcal{O} (\Gamma^2).
\label{eq:2_16b}
\end{align}
\end{subequations}
The first-order correction of the spin angular vertex $s^k$ also contains two separate vertices
\begin{align}
\mathcal{S}^{k,(1)}_{\omega}  &= \frac{n_\text{i} u^2}{V} \sum_{\bm{k}} \mathcal{G}^R_{\bm{k}} (\omega)
s^k \mathcal{G}^A_{\bm{k}} (\omega) \notag \\
&= S_2^{\mathcal{M}}(\omega) s^k  + A^{\mathcal{M}}(\omega) m^k ,
\label{eq:2_17}
\end{align}
where the symmetric coefficient $ S_2^{\mathcal{M}}(\omega)$ is
\begin{align}
 S_2^{\mathcal{M}}(\omega) = \frac{ \frac{1}{3} (\hbar \omega + \mu)^2 +\frac{2}{3}m^2 }{ (\hbar \omega +\mu)^2 +m^2}   + \mathcal{O} (\Gamma^2).
\label{eq:2_18}
\end{align}
The iterative structure of the ladder diagram can be written in a matrix form
\begin{align}
\left(\begin{array}{c} \mathcal{S}^{k}_{\omega}  \\  \mathcal{M}^{k}_{\omega}  \end{array}\right)
&=   \sum^\infty_{i=1}  \left(\begin{array}{cc}   S_2^{\mathcal{M}}(\omega) & A^{\mathcal{M}}(\omega)  \\  A^{\mathcal{M}}(\omega) &  S^{\mathcal{M}}(\omega) \end{array}\right)^{i}
\left(\begin{array}{c} s^k \\ m^k \\
\end{array}\right) ,
\label{eq:c-8}
\end{align}
where the entanglement between the spin magnetic vertex and spin angular vertex can be easily traced from the diagonal part of the iterative matrix. The corrected spin magnetic vertex is, thus,
\begin{align}
\mathcal{M}^{k} &= \frac{(\mu^2 +m^2)}{2(\mu^2 -m^2)^2}  \left[ 9 m \mu  s^k +  (6 \mu^2 + 3m^2)  m^k \right] .
\label{eq:2_20}
\end{align}
For massive Dirac electrons, the corrected spin magnetic vertex diverges at the band edge. The divergence can be justified by the fact that at the band edge all the scattering processes are degraded due to the vanishing of the states, as will be discussed in Sec.~\ref{Sec.3-4}.
Note that for massless Dirac electrons, the iterative matrix in Eq.~(\ref{eq:c-8}) is diagonalized, indicating that the spin magnetic and spin angular vertices are fully decoupled and the corrected spin magnetic vertex is simply $\mathcal{M}^{k}_{0} = 2 m^k$, consistent with a previous investigation on the massless Weyl fermion\cite{taguchi2016photovoltaic}.

Similarly,  we check the first-order correction on the velocity vertex:
\begin{align}
\Lambda^{i,(1)}_{\omega,\omega+\Omega}  &= \frac{n_\text{i} u^2}{V} \sum_{\bm{k}} \mathcal{G}^R_{\bm{k}} (\omega)v_i \mathcal{G}^A_{\bm{k}} (\omega+\Omega)  \notag \\
 &= S^{\Lambda} (\omega, \omega+\Omega) v_i + A^{\Lambda} (\omega, \omega+\Omega) v^s_i ,
\label{eq:2_21}
\end{align}
with symmetric $S^{\Lambda} $ and anti-symmetric $ A^{\Lambda} $ coefficients:
\begin{subequations}
\begin{align}
\label{eq:2_22a}
 S^{\Lambda} (\omega, \Omega) 
 &\simeq \frac{1}{3}   \left(  1 - i \frac{\hbar \Omega }{\Gamma(\omega)}    \right)  \frac{  (\hbar \omega +\mu)^2 -  m^2  }{ ( \hbar \omega + \mu)^2 +m^2},   \\
 A^{\Lambda} (\omega, \Omega) 
 &\simeq \left(  1 - i \frac{\hbar \Omega }{\Gamma(\omega)}    \right)  \frac{ 2 m \gamma_0 (\omega)  }{ ( \hbar \omega + \mu)^2 + m^2} .
\label{eq:2_22b}
\end{align}
\end{subequations}
Note that the reciprocal vertex $v^s_i \equiv  \rho_2 \otimes \sigma^i $ is often referred as spin velocity\cite{fujimoto2018transport}. Clearly, the antisymmetric coefficient of the iterative matrix for velocity is negligible compared with the symmetric one. Hence, the iterative matrix for the velocity vertex is approximately diagonalized, indicating the decoupling between the velocity vertex and spin velocity even in massive Dirac electrons. After summing the infinite ladder diagram, the corrected velocity vertex is
\begin{align}
\Lambda^{i}_{\Omega} 
& = \frac{\mu^2 - m^2}{2( \mu^2 +2m^2 )} \left( 1 -  i \frac{\hbar \Omega }{\Gamma}  \frac{3(\mu^2 + m^2)}{2( \mu^2 +2m^2 )}  \right) v_i .
\label{eq:2_23}
\end{align}
Clearly, the corrected velocity vertex only contains terms associated with $v_i $.

\begin{figure}[h]
 \includegraphics[width=8cm]{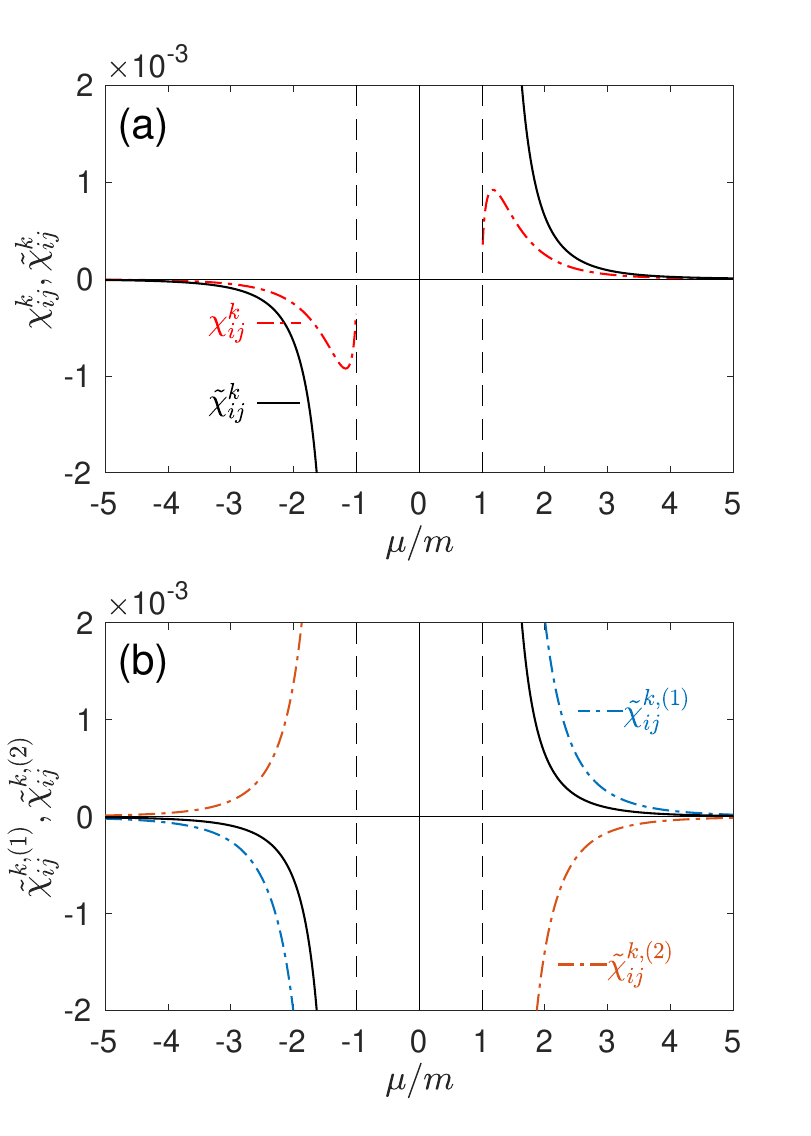}
 \caption{\label{fig:2} (a) The leading order $[\mathcal{O}(\Omega^1)]$ of spin magnetic response with $\tilde{\chi}^k_{ij}$ or without $\chi^k_{ij}$ vertex corrections. (b) Two contributions $\tilde{\chi}^{k,(1)}_{ij}, \tilde{\chi}^{k,(2)}_{ij}$ to the spin magnetic response with vertex correction. The light frequency is set as $ \Omega \tau_0 = 0.1$. The unit of the response coefficient is normalized by $\chi_0 = \hbar  e^2 v^2 \tau_0^2 \nu_0 $ with $\nu_0= m^2/(\pi^2 \hbar^3v^3)$ and $\tau_0 =\hbar/(\pi n_{\text{i}} u^2 \nu_0)$.}
 \end{figure}

Including the VCs on both the spin magnetic and spin velocity (see Fig.~\ref{fig:1}), the spin magnetic response function can be separated into two contributions as
\begin{align}
\tilde{\chi}^{k}_{ij} (\Omega) &=\tilde{\chi}^{k,(1)}_{ij} (\Omega)  + \tilde{\chi}^{k,(2)}_{ij} (\Omega) ,
\label{eq:2_24}
\end{align}
with
\begin{subequations}
\begin{align}
\label{eq:2_24(a)}
\tilde{\chi}^{k,(1)}_{ij} (\Omega)&\equiv  \left( \frac{ 3 (2 \mu^2 +m^2) (\mu^2 +m^2)}{2(\mu^2 -m^2)^2} +  \frac{ \mu^2 -m^2}{\mu^2 + 2 m^2}  \right) \chi^k_{ij} (\Omega) , \\
 \tilde{\chi}^{k,(2)}_{ij} (\Omega) &\equiv \frac{ 9 m \mu(\mu^2 +m^2)}{2(\mu^2 -m^2)^2}  \Pi^k_{ij} (\Omega) .
\label{eq:2_24(b)}
\end{align}
\end{subequations}
Due to mixing of the spin magnetic and spin angular vertices, the corrected spin magnetic response $\tilde{\chi}^{k}_{ij}$ contains two terms proportional to the spin magnetic, $\tilde{\chi}^{k,(1)}_{ij} (\Omega) \propto \chi^k_{ij} (\Omega)$, and spin angular responses $\tilde{\chi}^{k,(2)}_{ij} (\Omega) \propto  \Pi^k_{ij} (\Omega) $, respectively. In Fig.~\ref{fig:2} (b), we separately present the two contributions which show opposite sign and both diverge at the band edges. 
The total spin magnetic response $\tilde{\chi}^k_{ij} (\Omega)$ including the VCs still vanishes in the massless case, indicating the IFE only exists in massive Dirac electrons. The response coefficient is drastically enhanced in the vicinity of the band gap and shows divergent behavior at the band edge [Fig.~\ref{fig:2} (a)], attributed to the divergent behavior of the VC on the spin magnetic vertex [see Eq.~\ref{eq:2_20}]. Note that the present analysis breaks down only at the band edge due to a divergence of the electrons' lifetime.

 \subsection{Long-range diffusion}\label{Sec.3-4}
The divergence of VC indicates a long-range diffusion\cite{coleman2015introduction}. In fact,
considering the finite external wavevector $\bm{q}$ and frequency $\nu$, the VC of the spin
magnetic vertex (see Appendix.~\ref{Appx.B}) becomes
\begin{align}
\mathcal{M}^{k} (\bm{q}, \nu)  &\simeq \frac{1 }{\tilde{D} q^2  \tau  - i \nu \tilde{\tau} + K} \notag \\
&\times \left(  \frac{m\mu}{(m^2 +\mu^2) } s^k+\frac{m^2 +2 \mu^2}{3(m^2 +\mu^2) } m^k \right)
\label{eq:2_25}
\end{align}
with normalized diffusion coefficients $\tilde{D}$ and normalized lifetime $\tilde{\tau}$. The constant $K \equiv \frac{ 2 (\mu^2 - m^2 )^2 }{9 (\mu^2 +m^2 )^2 }$ is the \textit{mass} term of the diffusion kernel, which gives the spin diffusion length $\lambda_s $:
\begin{align}
\lambda_s &= \sqrt{  \tilde{D} \tau/ K}
\label{eq:2_26}
\end{align}
The spin diffusion length diverges as $\lambda_s \propto ( \mu^2 -m^2)^{-1}$ at the band edge for both the massive and massless cases. Note that for the massless Dirac Hamiltonian, the band edge returns to a single Dirac point $(\mu=0)$. For the massless case, the \textit{mass} term $K$ is a constant, wherefore the spin density profile drastically decays at the Dirac point without divergence. For the massive Dirac electron, the \textit{mass} term $K$, however, vanishes at the band edge, causing the divergence of the spin density profile.

\section{Discussion and Summary}
We theoretically study the spin magnetic density induced by the circularly polarized light in massive Dirac electrons with random short-ranged impurities. The induced spin magnetic density only appears in the presence of the mass gap and is linear with the photon energy up to the THz regime. The vertex correction drastically enhances the spin magnetic response in the vicinity of the band edge. 
 
In the Dirac Hamiltonian, the vertex correction due to short-ranged impurities on the spin magnetic vertex involves the mixing between the spin magnetic vertex and spin angular vertex  in the massive case. In general, the VCs of the Dirac Hamiltonian involves an entanglement between the original vertex, denoted by $O$, and its counterpart vertex obtained from the anti-commutator with the mass term operator, $\{O , \rho_3 \otimes \sigma^0\}$. In the case of the velocity vertex, the mixing is negligible in the order of $\Gamma/\mu$ and the summation of the iterative ladders converge trivially, while it is the same order as the
original vertex for the case of the spin and charge density vertex (see Appendix.~\ref{Appx.C}), resulting in
the enhancement at band edge.
The entanglement of the vertices in VCs is the consequence of the mixing between the positive and negative energy states in massive Dirac electrons. The mixing is significant where the chemical potential is close to the Dirac point. For example, the corrected spin magnetic vertex [Eq.~(\ref{eq:2_20})] has a contribution from the spin angular vertex with the order of $m/\mu$. and the VCs return to nearly a constants $\tilde{\chi}^{k}_{ij} \sim 4  \chi^k_{ij}$, when $\mu$ is far from the band edge.

For massive Dirac electrons, the occurrence of the IFE do not require breaking the inversion symmetry, in contrast to the massless Weyl semimetal\cite{taguchi2016photovoltaic}. One typical material which could be effectively described by the Dirac Hamiltonian is the $L$-point of bismuth and its alloys\cite{fuseya2015transport}. We estimate the induced spin magnetic field $- g^*  \mu_B /2 \braket{\bm{m}}$ by the circularly polarized light with parameters: $v=8.2 \times 10^5 $ m/s, $m=7.7 $ meV, $g^* \sim 1000$,  $\tau=4.2 \times 10^{-13} $ s. The chemical potential is chosen as $\mu=10$ meV near the band edge. For the monochromatic light source, we set the frequency to the THz regime: $\Omega =1$ THz with electric field strength $|\bm{\mathcal{E}}| = 3.1$ kV/m ($I=1.3 \times 10^4$ W/m$^2$)\cite{kawaguchi2020giant} and the induced effective magnetic field is $1.8 \times 10^{-9} $ T. Such a spin magnetic moment excited by circularly polarized light at the surface is proposed to be detectable through electrical measurement via the inverse spin Hall effect\cite{kawaguchi2020giant}.
 For a pulse laser whose electric field strength can rise to $|\bm{\mathcal{E}}| = 0.27$ GV/m ($I=10^{14}$ W/m$^2$)\cite{mangin2014engineered}, the induced effective magnetic field can reach $13.6$ T, possibly attributed to the large effective $g^*$ of $L$-point in bismuth\cite{PhysRevLett.115.216401}. 
 Our analytical calculation can not be applied to the frequency range of the visible light, due to the assumption in the expansion of photon energy $\Omega \tau \ll 1$. The extension to the visible light range could rely on the numerical calculations based on the density functional theory \cite{PhysRevB.94.144432,PhysRevLett.117.137203}.

\begin{acknowledgments}
The authors would like to thank Professor M. Hayashi and Dr. J Fujimoto for their kind correspondences. This work was supported by Japan Society for the Promotion of Science KAKENHI (21H01034).\end{acknowledgments}

\appendix
\section{ Calculation of spin magnetic response function}\label{Appx.A}

 \begin{widetext}
Expand the lesser component in Eq.~\ref{eq:2_9}, spin magnetic response function $\chi^k_{ij}( \Omega)$ reads
\begin{align}
\chi^k_{ij}( \Omega) &=  \frac{ e^2 }{4 \hbar\Omega^2 V} \sum_{\bm{k},\omega} \text{Tr} \Big[ 
\left(  f(\omega + \Omega ) -  f(\omega  ) \right) \Big(  m^k \mathcal{G}^R_{\bm{k}} (\omega )   v_i \mathcal{G}^A_{\bm{k}} (\omega + \Omega)  v_j   \mathcal{G}^A_{\bm{k}} (\omega ) -   m^k \mathcal{G}^R_{\bm{k}} (\omega )    v_i \mathcal{G}^R_{\bm{k}} (\omega + \Omega) v_j \mathcal{G}^A_{\bm{k}} (\omega) 
\Big) \notag\\
&+ f(\omega) \Big(  m^k  \mathcal{G}^A_{\bm{k}} (\omega)   v_i  \mathcal{G}^A_{\bm{k}} (\omega + \Omega)    v_j    \mathcal{G}^A_{\bm{k}} (\omega) 
-  m^k \mathcal{G}^R_{\bm{k}} (\omega)   v_i  \mathcal{G}^R_{\bm{k}} (\omega + \Omega)    v_j    \mathcal{G}^R_{\bm{k}} (\omega)   \Big)  \Big] .
\label{eq:b-5}
\end{align}

Assuming a small frequency of the external light $\Omega \tau \ll 1$, the Green's functions can be expanded in order of $\Omega$:
\begin{align}
 \mathcal{G}^A_{\bm{k}} ( \omega + \Omega)  &=  \mathcal{G}^A_{\bm{k}} ( \omega ) + \Omega \frac{d}{d\omega}
(\mathcal{G}^A_{\bm{k}} ( \omega))  + \frac{\Omega^2}{2} \frac{d^2}{d\omega^2}
(\mathcal{G}^A_{\bm{k}} ( \omega)) +  \frac{\Omega^3}{6} \frac{d^3}{d\omega^3}
(\mathcal{G}^A_{\bm{k}} ( \omega))   +\mathcal{O}(\Omega^4) \notag \\
&=  \mathcal{G}^A_{\bm{k}} ( \omega ) -  \hbar \Omega (  \mathcal{G}^{A}_{\bm{k}} ( \omega  ) )^2+   \hbar^2 \Omega^2   (  \mathcal{G}^{A}_{\bm{k}} (  \omega ) )^3 -   \hbar^3 \Omega^3   (  \mathcal{G}^{A}_{\bm{k}} ( \omega ) )^4  +\mathcal{O}(\Omega^4),
\label{eq:b-6}
\end{align}
where we recall identity about the partial derivatives of the  Green's functions: $d_{\omega}  \mathcal{G}^{R,A}_{\bm{k}} (\omega) = - \hbar  ( \mathcal{G}^{R,A}_{\bm{k}} (\omega) )^2 $.

Note that all even order of $\Omega$ terms vanishes. Hence, we only consider the odd order of $\Omega$ terms. The \textbf{first order} of $\Omega$-expansion in Eq.~\ref{eq:b-5} is the $\Omega^{-1}$ term:
\begin{align}
\chi^{k,(-1)}_{ij}( \Omega) &= \frac{e^2 }{4 \hbar \Omega V} \sum_{\bm{k},\omega} \text{Tr} \Big[ 
  f'(\omega  ) \Big(   m^k \mathcal{G}^R_{\bm{k}} (\omega)     v_i \mathcal{G}^A_{\bm{k}} (\omega )   v_j    \mathcal{G}^A_{\bm{k}} (\omega) -   m^k\mathcal{G}^R_{\bm{k}} (\omega)   v_i \mathcal{G}^R_{\bm{k}} (\omega ) v_j     \mathcal{G}^A_{\bm{k}} (\omega) 
\Big) \notag\\
&-  f(\omega) \Big(  m^k  \mathcal{G}^A_{\bm{k}} (\omega)   v_i ( \mathcal{G}^A_{\bm{k}} (\omega) )^2     v_j   \mathcal{G}^A_{\bm{k}} (\omega) 
- m^k \mathcal{G}^R_{\bm{k}} (\omega)    v_i  ( \mathcal{G}^R_{\bm{k}} (\omega) )^2    v_j   \mathcal{G}^R_{\bm{k}} (\omega)   \Big)  \Big] \notag \\
&= \frac{e^2 }{4 \hbar \Omega V} \sum_{\bm{k},\omega} \text{Tr} \Big[ 
  f'(\omega  ) \Big(   m^k \mathcal{G}^R_{\bm{k}} (\omega)     v_i \frac{ \partial \mathcal{G}^A_{\bm{k}} (\omega) }{\partial k_j}   -   m^k \frac{ \partial \mathcal{G}^R_{\bm{k}} (\omega) }{\partial k_i}  v_j     \mathcal{G}^A_{\bm{k}} (\omega) 
\Big) \notag\\
&- \hbar  f(\omega) \Big(  m^k  \frac{ \partial \mathcal{G}^A_{\bm{k}} (\omega) }{\partial k_i}  \frac{ \partial \mathcal{G}^A_{\bm{k}} (\omega) }{\partial k_j}
- m^k    \frac{ \partial \mathcal{G}^R_{\bm{k}} (\omega) }{\partial k_i}  \frac{ \partial \mathcal{G}^R_{\bm{k}} (\omega) }{\partial k_j} \Big)  \Big] .
\label{eq:b-9}
\end{align}
It is easily checked that the Fermi sea term $(\propto f(\omega))$ vanishes after taking the trace. By using the fact $\chi^{k,(-1)}_{ij}( \Omega) = \frac{1}{2} \left( \chi^{k,(-1)}_{ij}( \Omega) -\chi^{k,(-1)}_{ji}( \Omega)  \right)$, the Fermi sea term becomes
\begin{align}
\chi^{k,(-1)}_{ij}( \Omega) 
&= \frac{e^2 }{8 \hbar \Omega V} \sum_{\bm{k},\omega}   f'(\omega  )  \text{Tr} \Big[ 
\partial_{k_j}\Big(   m^k \mathcal{G}^R_{\bm{k}} (\omega)     v_i \mathcal{G}^A_{\bm{k}} (\omega) \Big)   - \partial_{k_i} \Big(   m^k  \mathcal{G}^R_{\bm{k}} (\omega)  v_j     \mathcal{G}^A_{\bm{k}} (\omega)  \Big) \Big]  \notag\\
&= \frac{  2 \hbar v^2 e^2 \gamma_3}{  \Omega (2\pi)^3} \sum_{\omega}   f'(\omega  )  \int d k_i dk_k  \frac{ \epsilon_{ijk} k_j }{D^R(\omega) D^A(\omega) }  \Big|^{k_j = + \infty}_{k_j = - \infty}  =0,
\label{eq:b-10}
\end{align}
where we notice the integral vanishes due to the cyclic symmetry of $k_1,k_2,k_3$ (spherical symmetry). Thus, the \textbf{first order} of $\Omega$ term vanishes totally, $\chi^{k,(-1)}_{ij}( \Omega)=0$.

The \textbf{third order} of $\Omega$-expansion in Eq.~\ref{eq:b-5} is the $\Omega^{1}$ term:
\begin{align}
\chi^{k,(1)}_{ij}( \Omega) &= \frac{e^2  \Omega }{4 \hbar V} \sum_{\bm{k},\omega} \text{Tr} \Big[ 
\hbar^2  f'(\omega  ) \Big(   m^k \mathcal{G}^R_{\bm{k}} (\omega)     v_i ( \mathcal{G}^A_{\bm{k}} (\omega ) )^3  v_j    \mathcal{G}^A_{\bm{k}} (\omega) -   m^k\mathcal{G}^R_{\bm{k}} (\omega)   v_i ( \mathcal{G}^R_{\bm{k}} (\omega ) )^3 v_j     \mathcal{G}^A_{\bm{k}} (\omega) 
\Big) \notag\\
&-\frac{1}{2} \hbar  f''(\omega  ) \Big(   m^k \mathcal{G}^R_{\bm{k}} (\omega)     v_i ( \mathcal{G}^A_{\bm{k}} (\omega ) )^2  v_j    \mathcal{G}^A_{\bm{k}} (\omega) -   m^k\mathcal{G}^R_{\bm{k}} (\omega)   v_i  ( \mathcal{G}^R_{\bm{k}} (\omega ) )^2  v_j     \mathcal{G}^A_{\bm{k}} (\omega) 
\Big) \notag\\
&+ \frac{1}{6} f'''(\omega  ) \Big(   m^k \mathcal{G}^R_{\bm{k}} (\omega)     v_i  \mathcal{G}^A_{\bm{k}} (\omega )   v_j    \mathcal{G}^A_{\bm{k}} (\omega) -   m^k\mathcal{G}^R_{\bm{k}} (\omega)   v_i   \mathcal{G}^R_{\bm{k}} (\omega )   v_j     \mathcal{G}^A_{\bm{k}} (\omega) 
\Big) \notag\\
& - \hbar^3 f(\omega) \Big(  m^k  \mathcal{G}^A_{\bm{k}} (\omega)   v_i ( \mathcal{G}^A_{\bm{k}} (\omega) )^4     v_j   \mathcal{G}^A_{\bm{k}} (\omega) 
- m^k \mathcal{G}^R_{\bm{k}} (\omega)    v_i  ( \mathcal{G}^R_{\bm{k}} (\omega) )^4    v_j   \mathcal{G}^R_{\bm{k}} (\omega)   \Big)  \Big] .
\label{eq:b-11}
\end{align}
The Fermi sea term reads
\begin{align}
 \frac{ \hbar^2 e^2  \Omega }{4  V} \text{Im} \sum_{\bm{k},\omega} f (\omega) \text{Tr} \Big[  m^k  \mathcal{G}^A_{\bm{k}} (\omega)   v_i ( \mathcal{G}^A_{\bm{k}} (\omega) )^4     v_j   \mathcal{G}^A_{\bm{k}} (\omega) \Big] 
&=   \frac{ 2 \hbar^2 e^2  \Omega }{  V}  \text{Re}  \sum_{\bm{k},\omega} f (\omega)   \epsilon_{ijk} \frac{   ( m + i \gamma_3) ( \hbar \omega + \mu - i \gamma_0) }{ ( D^A (\omega) )^4 }  =\mathcal{O}(\Gamma^1),
\label{eq:b-12}
\end{align}
which is higher order of $\Gamma$ comparing with the following Fermi surface term. 

For the fermi surface term, the third term $(\propto f'''(\omega))$ vanishes in the same way as Eq.~\ref{eq:b-10} and the rest terms can be rewritten as
\begin{align}
& \sum_{\bm{k},\omega} \text{Tr} \Big[ 
\hbar^2  f'(\omega  ) \Big(   m^k \mathcal{G}^R_{\bm{k}} (\omega)     v_i ( \mathcal{G}^A_{\bm{k}} (\omega ) )^3  v_j    \mathcal{G}^A_{\bm{k}} (\omega) -   m^k\mathcal{G}^R_{\bm{k}} (\omega)   v_i ( \mathcal{G}^R_{\bm{k}} (\omega ) )^3 v_j     \mathcal{G}^A_{\bm{k}} (\omega) 
\Big) \notag\\
&- \frac{1}{2} \hbar f''(\omega  ) \Big(   m^k \mathcal{G}^R_{\bm{k}} (\omega)     v_i ( \mathcal{G}^A_{\bm{k}} (\omega ) )^2  v_j    \mathcal{G}^A_{\bm{k}} (\omega) -   m^k\mathcal{G}^R_{\bm{k}} (\omega)   v_i  ( \mathcal{G}^R_{\bm{k}} (\omega ) )^2  v_j     \mathcal{G}^A_{\bm{k}} (\omega) 
\Big) \notag\\
 &=\sum_{\bm{k},\omega} \text{Tr} \Big[ 
\hbar^2  f'(\omega  ) \Big(   m^k \mathcal{G}^R_{\bm{k}} (\omega)     v_i ( \mathcal{G}^A_{\bm{k}} (\omega ) )^3  v_j    \mathcal{G}^A_{\bm{k}} (\omega) -   m^k\mathcal{G}^R_{\bm{k}} (\omega)   v_i ( \mathcal{G}^R_{\bm{k}} (\omega ) )^3 v_j     \mathcal{G}^A_{\bm{k}} (\omega) 
\Big) \notag\\
&+ \frac{1}{2} \hbar f'(\omega  )  \partial_{\omega} \Big(   m^k \mathcal{G}^R_{\bm{k}} (\omega)     v_i ( \mathcal{G}^A_{\bm{k}} (\omega ) )^2  v_j    \mathcal{G}^A_{\bm{k}} (\omega) -   m^k\mathcal{G}^R_{\bm{k}} (\omega)   v_i  ( \mathcal{G}^R_{\bm{k}} (\omega ) )^2  v_j     \mathcal{G}^A_{\bm{k}} (\omega) 
\Big) \notag\\
 &=\sum_{\bm{k},\omega} \text{Tr} \Big[ 
\hbar^2  f'(\omega  ) \Big(   m^k \mathcal{G}^R_{\bm{k}} (\omega)     v_i ( \mathcal{G}^A_{\bm{k}} (\omega ) )^3  v_j    \mathcal{G}^A_{\bm{k}} (\omega) -   m^k\mathcal{G}^R_{\bm{k}} (\omega)   v_i ( \mathcal{G}^R_{\bm{k}} (\omega ) )^3 v_j     \mathcal{G}^A_{\bm{k}} (\omega) 
\Big) \notag\\
&- \frac{1}{2} \hbar^2 f'(\omega  )   \Big(   m^k ( \mathcal{G}^R_{\bm{k}} (\omega)  )^2   v_i ( \mathcal{G}^A_{\bm{k}} (\omega ) )^2  v_j    \mathcal{G}^A_{\bm{k}} (\omega)
+ 2 m^k  \mathcal{G}^R_{\bm{k}} (\omega)     v_i ( \mathcal{G}^A_{\bm{k}} (\omega ) )^3  v_j    \mathcal{G}^A_{\bm{k}} (\omega) + m^k  \mathcal{G}^R_{\bm{k}} (\omega)     v_i ( \mathcal{G}^A_{\bm{k}} (\omega ) )^2  v_j   ( \mathcal{G}^A_{\bm{k}} (\omega) )^2 \notag \\
& -  m^k ( \mathcal{G}^R_{\bm{k}} (\omega)  )^2   v_i ( \mathcal{G}^R_{\bm{k}} (\omega ) )^2  v_j    \mathcal{G}^A_{\bm{k}} (\omega)
- 2 m^k  \mathcal{G}^R_{\bm{k}} (\omega)     v_i ( \mathcal{G}^R_{\bm{k}} (\omega ) )^3  v_j    \mathcal{G}^A_{\bm{k}} (\omega) - m^k  \mathcal{G}^R_{\bm{k}} (\omega)     v_i ( \mathcal{G}^R_{\bm{k}} (\omega ) )^2  v_j   ( \mathcal{G}^A_{\bm{k}} (\omega) )^2 
\Big) \notag\\
 &=\sum_{\bm{k},\omega} - \frac{1}{2} \hbar^2 f'(\omega  ) \text{Tr} \Big[   m^k ( \mathcal{G}^R_{\bm{k}} (\omega)  )^2   v_i ( \mathcal{G}^A_{\bm{k}} (\omega ) )^2  v_j    \mathcal{G}^A_{\bm{k}} (\omega)
+ m^k  \mathcal{G}^R_{\bm{k}} (\omega)     v_i ( \mathcal{G}^A_{\bm{k}} (\omega ) )^2  v_j   ( \mathcal{G}^A_{\bm{k}} (\omega) )^2 \notag \\
& -  m^k ( \mathcal{G}^R_{\bm{k}} (\omega)  )^2   v_i ( \mathcal{G}^R_{\bm{k}} (\omega ) )^2  v_j    \mathcal{G}^A_{\bm{k}} (\omega) - m^k  \mathcal{G}^R_{\bm{k}} (\omega)     v_i ( \mathcal{G}^R_{\bm{k}} (\omega ) )^2  v_j   ( \mathcal{G}^A_{\bm{k}} (\omega) )^2 
\Big],
\label{eq:b-13}
\end{align}
with which $\chi^{k,(1)}_{ij}( \Omega)$  gives Eq.~\ref{eq:2_12}.
 
At zero temperature limit, the derivative of Fermi distribution returns to the $\delta$-function $\omega=0$:
\begin{align}
\chi^{k,(1)}_{ij}( \Omega) &=  \frac{ \hbar e^2  \Omega }{16 \pi  V}  \sum_{\bm{k}} \text{Tr} \Big[ 
  m^k ( \mathcal{G}^R_{\bm{k}} ( 0 )  )^2   v_i ( \mathcal{G}^A_{\bm{k}} ( 0  ) )^2  v_j    \mathcal{G}^A_{\bm{k}} ( 0 )  - m^k  \mathcal{G}^R_{\bm{k}} ( 0 )     v_i ( \mathcal{G}^R_{\bm{k}} ( 0 ) )^2  v_j   ( \mathcal{G}^A_{\bm{k}} ( 0 ) )^2  \notag \\
&+ m^k  \mathcal{G}^R_{\bm{k}} ( 0 )     v_i ( \mathcal{G}^A_{\bm{k}} ( 0 ) )^2  v_j   ( \mathcal{G}^A_{\bm{k}} ( 0 ) )^2
 -  m^k ( \mathcal{G}^R_{\bm{k}} ( 0 )  )^2   v_i ( \mathcal{G}^R_{\bm{k}} ( 0 ) )^2  v_j    \mathcal{G}^A_{\bm{k}} ( 0 )\Big] .
\label{eq:b-14}
\end{align}

The first term $(g^Rg^Rg^Ag^Ag^A-g^Rg^Rg^Rg^Ag^A)$ in Eq.~\ref{eq:b-14} is
\begin{align}
&\frac{ \hbar e^2  \Omega }{16 \pi  V}\sum_{\bm{k}} \frac{1}{ ( D^R D^A )^2} \left( \frac{ \text{Tr} \Big[    m^k (g^R)^2 v_i  (g^A)^2   v_j g^A \Big] }{  D^A }  - \frac{ \text{Tr} \Big[    m^k g^R v_i  (g^R)^2   v_j ( g^A )^2\Big] }{ D^R }  \right) \notag \\
&= \frac{ \hbar e^2  \Omega }{8 \pi  V}\sum_{\bm{k}} \text{Re} \left( \frac{ \text{Tr} \Big[    m^k (g^R)^2 v_i  (g^A)^2   v_j g^A \Big] }{ ( D^R D^A )^2  D^A }   \right) \notag \\
&= \epsilon_{ijk} \hbar^3 v^2 e^2   \frac{ m (2\mu^2 -m^2)  }{ 32   \mu^3 (\mu^2 +m^2)  \Gamma^2 }   \nu(\mu) \Omega +\mathcal{O}(\Gamma^0),
\label{eq:b-15}
\end{align}
where $D^{R,A}$ and $g^{R,A}$ are short for $D^{R,A} (0)$ $g^{R,A}(0)$ for the denominator and numerator of the Green's function at chemical potential $\mu$. 

Note that it is easily checked that
\begin{align}
\left( \text{Tr} \Big[    m^k (g^R)^2 v_i  (g^A)^2   v_j g^A \Big] \right)^* = - \text{Tr} \Big[   m^k   g^R  v_i   (g^R)^2 v_j  (g^A)^2    \Big] 
\label{eq:b-16}
\end{align}
where in the last equity we employ the fact the trace term changes sign when exchanging indices $i,j$.

The second term $(g^Rg^Ag^Ag^Ag^A-g^Rg^Rg^Rg^Rg^A)$ in Eq.~\ref{eq:b-14}  is
\begin{align}
&\frac{ \hbar e^2  \Omega }{16 \pi  V} \sum_{\bm{k}} \frac{1}{ D^R D^A}  \left(  \frac{ \text{Tr} \Big[    m^k g^R v_i  (g^A)^2   v_j (g^A)^2 \Big] }{ ( D^A )^3}  - \frac{ \text{Tr} \Big[    m^k ( g^R )^2 v_i  (g^R)^2   v_j  g^A \Big] }{( D^R )^3}\right) \notag \\
&=\frac{ \hbar e^2  \Omega }{ 8 \pi  V} \sum_{\bm{k}} \text{Re} \left(  \frac{ \text{Tr} \Big[    m^k g^R v_i  (g^A)^2   v_j (g^A)^2 \Big] }{D^R D^A ( D^A )^3}  \right)  \notag \\
&=  - \epsilon_{ijk}  \hbar^3 v^2 e^2     \frac{ m (2 \mu^4 + \mu^2 m^2+3 m^2)  }{96 \mu^3 (\mu^2 +m^2)^2\Gamma^2 }\nu(\mu) \Omega +\mathcal{O}(\Gamma^0).
\label{eq:b-17}
\end{align}

The sum over two terms gives Eq.~\ref{eq:2_13}:
\begin{align}
\chi^{k,(1)}_{ij}( \Omega)  &= \epsilon_{ijk} \hbar^3 v^2 e^2     \frac{ m (\mu^2 -m^2)(2\mu^2 +3 m^2)  }{48 \mu^3 (\mu^2 +m^2)^2\Gamma^2 }\nu(\mu) \Omega \notag \\
&=  \epsilon_{ijk}   \hbar v^2 e^2    \frac{    m  ( \mu^2 -m^2) ( 2\mu^2 +3 m^2) }{12  \mu^3 (\mu^2+m^2)^2 }  \nu(\mu) \Omega \tau^2.
\label{eq:b-18}
\end{align}
 \end{widetext}

\section{ spin magnetic vertex correction with finite $\bm{q},\nu$}\label{Appx.B}
For the spatial and time variation of the spin density, we need to consider vertex correction on the spin magnetic vertex with finite momentum and frequency $\bm{q},\nu$. The vertex correction also involves a summation over iterative matrix as Eq.~\ref{eq:c-8}. We consider first order correction of spin magnetic vertex:
\begin{align}
\mathcal{M}^{k,(1)} (\bm{q}, \nu) &= \frac{n_\text{i} u^2}{V} \sum_{\bm{k}} \mathcal{G}^R_{\bm{k}+ \bm{q}/2} \left( \frac{\nu}{2} \right) m^k \mathcal{G}^A_{\bm{k}-\bm{q}/2} \left( -\frac{\nu}{2} \right). \label{eq:A-1}
\end{align}

The integrand in Eq.~\ref{eq:A-1} is expanded to the second order of $q_i$ and first order of $\nu$:
\begin{align}
&\mathcal{G}^R_{\bm{k}+ \bm{q}/2} (\frac{\nu}{2}) m^k \mathcal{G}^A_{\bm{k}-\bm{q}/2} (-\frac{\nu}{2}) = \mathcal{G}^R_{\bm{k}} (0) m^k  \mathcal{G}^A_{\bm{k}} (0) \notag \\
&+  \frac{\nu}{2} \left( \partial_{\nu}  \mathcal{G}^R_{\bm{k}} (0) m^k  \mathcal{G}^A_{\bm{k}} (0) - \mathcal{G}^R_{\bm{k}} (0) m^k  \partial_{\nu}  \mathcal{G}^A_{\bm{k}} (0) \right) \notag \\
&+  \frac{q_i}{2} \left( \partial_{i}  \mathcal{G}^R_{\bm{k}} (0) m^k \mathcal{G}^A_{\bm{k}} (0) - \mathcal{G}^R_{\bm{k}} (0) m^k  \partial_{i}  \mathcal{G}^A_{\bm{k}} (0) \right) \notag \\
 &- \frac{\nu q_i}{4} \left( \partial_{\nu}  \mathcal{G}^R_{\bm{k}} (0) m^k \partial_{i}  \mathcal{G}^A_{\bm{k}} (0) +  \partial_{i}  \mathcal{G}^R_{\bm{k}} (0) m^k \partial_{\nu}  \mathcal{G}^A_{\bm{k}} (0)  \right) \notag \\
 &+ \frac{q_i q_j}{8} \partial_i  \partial_j  \mathcal{G}^R_{\bm{k}} (0) m^k  \mathcal{G}^A_{\bm{k}} (0) + \frac{q_i q_j}{8}   \mathcal{G}^R_{\bm{k}} (0) m^k \partial_i  \partial_j  \mathcal{G}^A_{\bm{k}} (0)\notag \\
&  - \frac{q_i q_j}{4}  \partial_{i}  \mathcal{G}^R_{\bm{k}} (0) m^k  \partial_{j}  \mathcal{G}^A_{\bm{k}} (0) +\mathcal{O} (q^3,\nu^2).
\label{eq:A-2}
\end{align}

The zeroth order correction is already shown in Sec.~\ref{Sec.3-3}. We start from the the  $\mathcal{O} (\nu^1)$ term:
\begin{align}
&\frac{n_\text{i} u^2}{V} \sum_{\bm{k}}   \frac{\nu}{2} \left( \partial_{\nu}  \mathcal{G}^R_{\bm{k}} (0) m^k  \mathcal{G}^A_{\bm{k}} (0) - \mathcal{G}^R_{\bm{k}} (0) m^k  \partial_{\nu}  \mathcal{G}^A_{\bm{k}} (0) \right) \notag \\
 &= - \frac{ i \hbar \nu n_\text{i} u^2}{V} \text{Im}  \sum_{\bm{k}}     \mathcal{G}^R_{\bm{k}} (0)  \mathcal{G}^R_{\bm{k}} (0) m^k  \mathcal{G}^A_{\bm{k}} (0) . 
\label{eq:A-3}
\end{align}

The $\mathcal{O} (\bm{q})$ term is
\begin{align}
&\frac{ i q_i n_\text{i} u^2}{2V}  \left(  \mathcal{G}^R_{\bm{k}} (0) v_i \mathcal{G}^R_{\bm{k}} (0) m^k \mathcal{G}^A_{\bm{k}} (0) - \mathcal{G}^R_{\bm{k}} (0) m^k \mathcal{G}^A_{\bm{k}} (0) v_i \mathcal{G}^A_{\bm{k}} (0) \right) \notag \\
& = \frac{ i q_i n_\text{i} u^2}{2V} \text{Im} \sum_{\bm{k}}      \mathcal{G}^R_{\bm{k}} (0) v_i \mathcal{G}^R_{\bm{k}} (0) m^k \mathcal{G}^A_{\bm{k}} (0)  = 0 .
\label{eq:A-4}
\end{align}
which vanishes due to the wave vector $k_i$ appears singly in the $\bm{k}$-integration.

The cross term $\mathcal{O} (\bm{q}\nu)$ also vanishes, due to the odd term of $k_i$ in the $\bm{k}$-integration.
\begin{align}
\frac{n_\text{i} u^2}{V}   \frac{ \hbar \nu q_i}{2} \text{Re}  \sum_{\bm{k}}     \mathcal{G}^R_{\bm{k}} (0)   \mathcal{G}^R_{\bm{k}} (0) m^k  \mathcal{G}^A_{\bm{k}} (0) v_i  \mathcal{G}^A_{\bm{k}} (0) =0 .
\label{eq:A-5}
\end{align}

In the second order $\mathcal{O} (q_i q_j)$, the $\bm{k}$-integral vanishes for $i \neq j$, since $k_i k_j$ terms appear in integration. For $i=j$ case, it reads
\begin{align}
& \frac{n_\text{i} u^2}{V}   \frac{ q^2  }{4} \text{Re}  \sum_{\bm{k}}   \partial^2_1  \mathcal{G}^R_{\bm{k}} (0) m^k  \mathcal{G}^A_{\bm{k}} (0) -  \partial_{1}  \mathcal{G}^R_{\bm{k}} (0) m^k  \partial_{1}  \mathcal{G}^A_{\bm{k}} (0)  \notag \\
&=  \frac{n_\text{i} u^2}{V}   \frac{ q^2  }{4} \text{Re}  \sum_{\bm{k}}   \mathcal{G}^R_{\bm{k}} (0) v_1 \mathcal{G}^R_{\bm{k}} (0)  v_1 \mathcal{G}^R_{\bm{k}} (0) m^k  \mathcal{G}^A_{\bm{k}} (0) \notag \\
& -  \mathcal{G}^R_{\bm{k}} (0) v_1  \mathcal{G}^R_{\bm{k}} (0) m^k  \mathcal{G}^A_{\bm{k}} (0)  v_1  \mathcal{G}^A_{\bm{k}} (0)  ,
\label{eq:A-6}
\end{align}
where $q=|\bm{q}|$. Note that we choose $i=j=1$ for the cyclic symmetry.

Evaluating the integration of $\mathcal{O} (\nu^1)$  and $\mathcal{O} (q_i q_i)$ terms, the iterative matrix $M_s$ reads
\begin{align}
M_s =\left(\begin{array}{cc} S_2^\mathcal{M} & A^\mathcal{M} \\ A^\mathcal{M} & S^\mathcal{M} \end{array}\right) ,
\label{eq:A-7}
\end{align}
with matrix elements are
\begin{subequations}
\begin{align}
\label{eq:A-8a}
S^\mathcal{M} & =  (1+  i \nu \tau) \frac{ 2 \mu^2 + m^2 }{3 \mu^2 +3 m^2 } - Dq^2 \tau \frac{11\mu^2  + 4m^2 }{ 18  (\mu^2 +m^2)} , \\
\label{eq:A-8b}
A^\mathcal{M} &  = (1+  i \nu \tau) \frac{  \mu m  }{ \mu^2 + m^2 }  - D q^2  \tau \frac{ 5 m \mu }{ 6  (\mu^2 +m^2)} ,  \\
\label{eq:A-8c}
S_2^\mathcal{M} & =(1+ i \nu \tau ) \frac{  \mu^2 +2 m^2 }{3 \mu^2 +3 m^2 }  - Dq^2  \tau \frac{ 4\mu^2 + 11 m^2}{ 18  (\mu^2 +m^2)},
\end{align}
\end{subequations}
where we define diffusion constant $D \equiv    \frac{  ( \mu^2 -m^2 ) v^2 \tau }{ 4 \mu^2}  $. Apparently, Eq.~\ref{eq:A-7} returns to Eq.~\ref{eq:c-8} for $\bm{q},\nu=0$.  Note that the diffusion constant vanishes at the band edge.
\begin{align}
D =    \frac{  ( \mu^2 -m^2 ) v^2 \tau  }{ 4 \mu^2}  \sim   \frac{  v^2   \sqrt{\mu^2- m^2} }{ 4 |\mu| (\mu^2 +m^2)} .
\label{eq:A-9}
\end{align}
For the massless case, the diffusion constant diverges at the Dirac point $\mu=0$.

Summing over the infinite ladder diagram, the power series reads
\begin{align}
\sum_{i=0}^{\infty} M_s^i = (1-M_s)^{-1}
\label{eq:A-10}
\end{align}
where the convergence of series requires $\mathrm{det}(1-M_s)\neq0$. Note that the trace of $M_s$ is
\begin{align}
\text{Tr} M_s = 1 - ( \frac{5}{6} D q^2  - i \nu \tau ).
\label{eq:A-11}
\end{align}
which always small than $1$ with finite $\bm{q}$. With two positive eigenvalues both smaller than $1$, the summation is always converged. Hence, the corrected spin magnetic vertex with finite $\bm{q},\nu$ is approximately

\begin{align}
\mathcal{M}^{k} (\bm{q}, \nu)  &\simeq \frac{1}{\tilde{D} q^2  \tau  - i \nu \tilde{\tau} + K} \notag \\
&\times \left(  \frac{m\mu}{(m^2 +\mu^2) } s^k+\frac{m^2 +2 \mu^2}{3(m^2 +\mu^2) } m^k  \right),
\label{eq:A-12}
\end{align}
with normalized diffusion coefficients:
\begin{subequations}
\begin{align}
\label{eq:A-13a}
\tilde{D} &\equiv   \frac{ ( 13 \mu^4  + 64 \mu^2 m^2 + 13 m^4 ) }{ 27 (\mu^2 +m^2 )^2 } D ,  \\
\label{eq:A-13b}
\tilde{\tau}  &\equiv \frac{ (\mu^2 + 5 m^2 ) (5\mu^2 +  m^2 ) }{9 (\mu^2 +m^2 )^2 } \tau ,\\
\label{eq:A-13c}
K &\equiv \frac{ 2 (\mu^2 - m^2 )^2 }{9 (\mu^2 +m^2 )^2 }.
\end{align}
\end{subequations}
The spin diffusion length is
\begin{align}
\lambda_s &= \sqrt{  \frac{\tilde{D} \tau}{K} } \notag \\
&= \sqrt{  \frac{  ( 13 \mu^4  + 64 \mu^2 m^2 + 13 m^4 ) }{  24 \mu^2 (\mu^2 - m^2 ) }    } v \tau  \notag \\
&\propto (\mu^2 - m^2 )^{-1},
\label{eq:A-14}
\end{align}
which is diverging at the band edge, similar with electrons' lifetime.

At the band edge, the \textit{mass} term $K$ for the diffusion term vanishes. Thus, the corrected spin magnetic vertex diverges in the $\bm{q},\nu=0$ limit. For the massless Dirac electron $(m=0)$, the  \textit{mass} term $K$ is a constant despite the position of chemical potential. Thus, the corrected spin magnetic vertex converges even in the $\bm{q},\nu=0$ limit.

\section{ charge vertex correction with finite $\bm{q},\nu$}\label{Appx.C}
Here, we consider the impurity vertex correction on the charge density $c_0 \equiv \rho_0 \otimes \sigma^0$. The first order correction of the charge density expanded with small $\bm{q},\nu$ is
\begin{align}
\mathcal{C}_0^{(1)} (\bm{q}, \nu) &= \frac{n_i u^2}{V} \sum_{\bm{k}} \mathcal{G}^R_{\bm{k}+ \bm{q}/2} \left( \frac{\nu}{2}\right) c_0 \mathcal{G}^A_{\bm{k}-\bm{q}/2}  \left(- \frac{\nu}{2}\right) ,
\label{eq:B-1}
\end{align}
where $c_3 \equiv \rho_3 \otimes \sigma^0 $ is the reciprocal operator of charge density. Note that the corrected $c_0$-vertex can be decomposed into $c_0$ and $c_3$ vertices for the massive case. Similarly, the corrected $c_3$-vertex can be also decomposed into $c_0$ and $c_3$.

In static limit $\bm{q},\nu=0$, the iterative matrix $M_c$ is 
\begin{align}
M_c = \frac{1}{\mu^2 +m^2} \left(\begin{array}{cc} \mu^2 & m \mu  \\ m \mu & m^2 \end{array}\right),
\label{eq:B-2}
\end{align}
in which we found $M^2=M$. Thus, the power series of $M$ matrix is
\begin{align}
\sum_{i}  M^i = 1+ M +M +M +M \cdots ,
\label{eq:B-3}
\end{align}
which is diverged despite of the massless case $(m=0)$ or at the band edge $(\mu = \pm m)$. Thus, the vertex correction is diffusive. It might be solved in higher order expansion of $\bm{q}$ and $\nu$.

Summing the zeroth order $\mathcal{O} (\bm{q}^0,\nu^0)$, first order of energy $\mathcal{O} (\bm{q}^0,\nu^1)$, and second order of momentum $\mathcal{O} (q^2,\nu^0)$ terms, the iterative matrix $M_c$ is 
\begin{align}
M_c &=  \frac{1 - (  D q^2 \tau- i\nu \tau)}{\mu^2 +m^2} \left(\begin{array}{cc} \mu^2 & m \mu  \\ m  \mu & m^2 \end{array}\right) ,
\label{eq:B-4}
\end{align}
where diffusion constant is denoted as $D \equiv - \frac{ v^2 q^2 \tau}{4}  \frac{  \mu^2 -m^2  }{\mu^2}$. Due to the existence of diffusion term, the summation converges despite chemical potential approaches the band edges $\mu= \pm m$. The corrected charge density is
\begin{align}
&\mathcal{C}_0 = \frac{1}{D q^2 -i \nu \tau} \notag \\
&\times \left( \frac{\mu^2 + m^2 (D q^2 -i \nu \tau )}{\mu^2 +m^2 } c_0 -  \frac{ m \mu  (1- D q^2 +i \nu \tau ) }{\mu^2 +m^2 } c_3 \right) .
\label{eq:B-5}
\end{align}

%


\bibliography{Dirac_IFE}

\end{document}